# Giant two-photon absorption in monolayer $MoS_2$

*Yuanxin Li[1], Ningning Dong[1], Saifeng Zhang[1], Xiaoyan Zhang[1], Yanyan Feng[1], Kangpeng Wang[1], Long Zhang[1], Jun Wang[1, 2, *]*

*Corresponding Author: jwang@siom.ac.cn

1 Key Laboratory of Materials for High-Power Laser, Shanghai Institute of Optics and Fine Mechanics, Chinese Academy of Sciences, Shanghai 201800, China.
2 State Key Laboratory of High Field Laser Physics, Shanghai Institute of Optics and Fine Mechanics, Chinese Academy of Sciences, Shanghai 201800, China

**Abstract** Strong two-photon absorption (TPA) in monolayer $MoS_2$ is demonstrated in contrast to saturable absorption (SA) in multilayer $MoS_2$ under the excitation of femtosecond laser pulses in the near infrared region. $MoS_2$ in the forms of monolayer single crystal and multilayer triangular islands are grown on either quartz or $SiO_2$/Si by employing the seeding method through chemistry vapor deposition. The nonlinear transmission measurements reveal that monolayer $MoS_2$ possesses a nonsaturation TPA coefficient as high as ∼$(7.62 \pm 0.15) \times 10^3$ cm/GW, larger than that of conventional semiconductors by a factor of $10^3$. As a result of TPA, two-photon pumped frequency up-converted luminescence is observed directly in the monolayer $MoS_2$. For the multilayer $MoS_2$, the SA response is demonstrated with the ratio of the excited-state absorption cross section to ground-state cross section of ∼0.18. In addition, the laser damage threshold of the monolayer $MoS_2$ is ∼97 $GW/cm^2$, larger than that of the multilayer $MoS_2$ of ∼78 $GW/cm^2$.

## 1. Introduction

Study of nonlinear absorption in MoS$_2$ and other layered transition metal dichalcogenides (TMDCs) has given rise to a new category of photonic nanomaterials for optical switching, ultrashort pulse generation, optical limiting, etc. [1-3]. MoS$_2$ in various forms prepared via, such as, pulsed laser deposition, chemical vapor deposition (CVD), liquid-exfoliation followed by polymer solidification, have been demonstrated successfully as a saturable absorber for mode-locking or Q-switching in ultrafast lasers over a broad wavelength range (1, 1.5 and 2 μm) [2,4-6]. However, the nonlinear absorption difference between MoS$_2$ monolayer and few-layer in the near infrared (NIR) remains unclear. In this paper, we reveal that monolayer MoS$_2$ possesses strong two-photon absorption (TPA) for femtosecond (fs) pulses at 1 μm and should not be suitable for mode-locking or Q-switching operation. By contrast, few-layer MoS$_2$ exhibits significant saturable absorption (SA) at the same wavelength, consistent with the reported results.

MoS$_2$ has a layered structure with a single layer of Mo atoms sandwiched by two layers of S atoms in the $D_{6h}^4$ crystal symmetry. When thinned down to monolayer, the crystal structure reduces to broken inversion symmetry $D_{3h}^1$, and MoS$_2$ changes from an indirect bandgap semiconductor with an energy gap of ~1.29 eV to a direct bandgap semiconductor reaching an optical bandgap of ~1.88 eV [7-14]. Because of this broken inversion symmetry as well as the spin-orbit coupling effect, the valence-band maximum of the monolayer at the *K* point is splitted. Taking the strong excitonic binding energy into consideration, the interband transitions energy are ~1.9 and ~2.0 eV, corresponding to the so-called A and B excitons, respectively [12-14].

With this transformation, the nonlinear optical (NLO) parametric process in MoS$_2$ exhibits an obvious layer-dependent behavior, e.g., the second-harmonic generation only exits in the odd-number layered MoS$_2$ [15,16], whereas third-harmonic generation can be used to identify MoS$_2$ atomic layers [17]. As for the NLO non-parametric process, such as nonlinear absorption, I.

Paradisanos etc. have reported that TPA and one photon absorption play important roles in the optical damage of monolayer and bulk MoS$_2$ [18]. However, how these nonlinear absorption in monolayer and multilayer MoS$_2$ perform remains unclear, and the systematic study as well as theory calculating of the nonlinear absorption parameters are significant for in-depth understanding of the optical non-parametric process and mechanism in pristine monolayer and multilayer MoS$_2$, which would be helpful for the application of this two dimensional (2D) semiconductor in photonics.

In this work, a comparative nonlinear absorption study on MoS$_2$ monolayer versus multilayer was conducted. The multilayer MoS$_2$ showed SA effect, whereas the monolayer MoS$_2$ exhibited TPA response for fs pulses in NIR. The TPA effect of monolayer MoS$_2$ was directly observed with a nonsaturation TPA coefficient of ~(7.62 $\pm$ 0.15) × 10$^3$ cm/GW, which is considerably larger than the other semiconductors, such as GaAs, GdS and ZnO [19-21]. We have reason to believe that the large TPA coefficient could be originated from the exciton enhancement due to the strong exciton-binding energy in monolayer MoS$_2$ [22, 23]. Profiting from this extraordinary TPA effect, two-photon pumped frequency up-converted luminescence was also observed. The results imply the importance of layer number engineering for the development of MoS$_2$, as well as the other TMDCs-based photonic devices, namely, passive mode-lockers, Q-switchers, optical limiters, light emitters, etc.

## 2. Results and discussion
### 2.1. The preparation and characterization of MoS$_2$

MoS$_2$ atomic layers were synthesized on 300 nm SiO$_2$/Si and transparent quartz, respectively, by CVD growth with PTAS seeding promoters (see the Supporting Information) [24-26]. Large scale continuous MoS$_2$ films with few square centimeters can be obtained on these two kinds of substrates, as shown in Figure 1 (a) and (b). Figure 1 (c) shows an optical microscopy (OM) image of the MoS$_2$ film on SiO$_2$/Si. Isolated islands with different thicknesses (monolayer and multilayer) were observed at the edge of the continuous film [Inset of Figure 1 (c)]. The edge lengths of the islands varied from several micrometers to tens of micrometers. A typical triangular monolayer MoS$_2$ island was shown in Figure 1 (d). Notably, similar results were acquired on the quartz plates, which are significant to perform the NLO measurement, inasmuch as it no longer need to transfer the MoS$_2$ film from SiO$_2$/Si to a transparent substrate. More importantly, this can avoid the damage caused by transferring and retain their intrinsic characteristics.

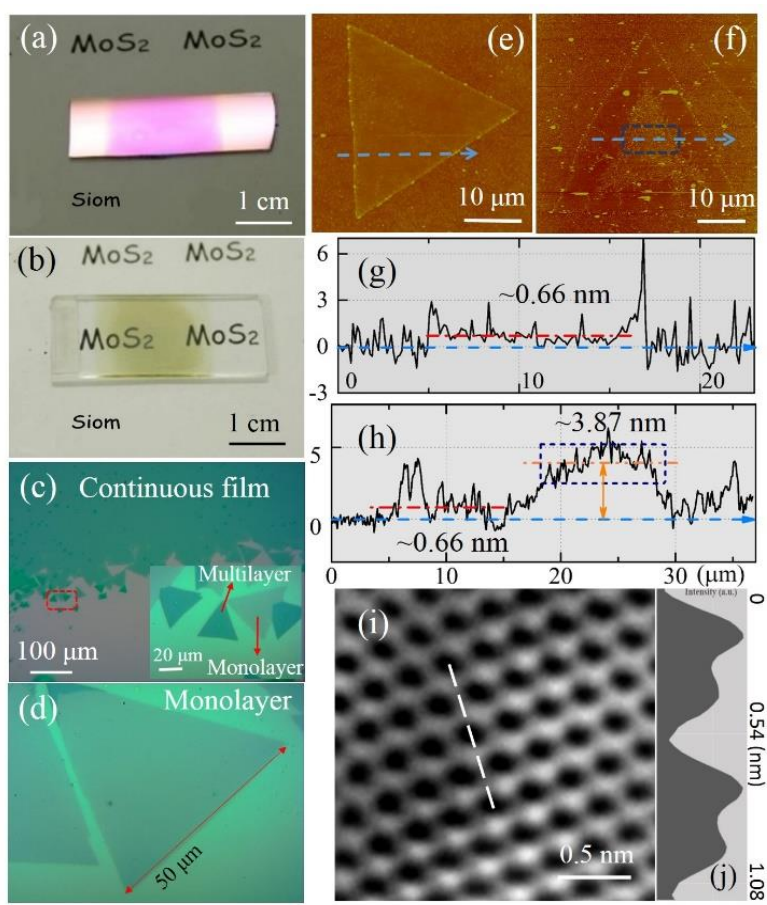

**Figure 1.** Photographs of MoS$_2$ grown on (a) SiO$_2$/Si, and (b) quartz. Optical microscopy images of (c) continuous MoS$_2$ film (10×); (inset) MoS$_2$ triangular islands with different thickness (100×) and (d) monolayer MoS$_2$ single crystal (100×) on SiO$_2$/Si. AFM topography of (e) monolayer and (f) multilayer MoS$_2$ on quartz with height profiles shown in (g) and (h), respectively. The height difference between the blue base line and the red line is ~0.66 nm in the two charts, indicating a monolayer. The yellow arrow shows the average height of ~3.87

nm, corresponding to 4 to 6 layers in the rectangular region. (i) HRTEM image of the transferred CVD $MoS_2$ with the (j) intensity profile measured along the white dot line.

To identify the height topography of these $MoS_2$ triangular islands, atomic force microscopy (AFM) measurements were carried out. Figure 1 (e)-(f) show the AFM images, and (g)-(h) are the corresponding height profiles. The homogeneous color contrast in Figure 1 (e) demonstrates that this island has a smooth surface with the thickness of ~0.66 nm [see Figure 1 (g)], which indicates that this $MoS_2$ triangular nanosheet is a monolayer. The little embossment at the edge was originated from the $MoS_2$ accumulation during the growth process. Figure 1 (f) shows a typical AFM image of multilayer $MoS_2$ island with non-uniform surface. The edge regions are clearly monolayer, whereas the centers are 4 to 6 layers, which can be confirmed from the height profile of Figure 1(h). This peculiar structure derives from the fact that the growth of $MoS_2$ favors layer growth at the initial nucleation site with complex subsequent growth and coalescence process [25, 27, 28]. To investigate the crystal structure of these islands, the high resolution transmission electron microscopy (HRTEM) measurement was carried out, and the HRTEM samples were prepared using a PMMA-assisted transfer technology [27]. Figure 1 (i) shows the HRTEM image preprocessed with digital periodic filter [3, 29], and the graphene-like hexagonal structure (2H) was clearly observed. By analyzing the intensity profiles along the white dot line in Figure 1 (i), we see a significant variation in intensity between neighboring atoms [see Figure 1 (j)]. Because of the ABAB stacking sequence of the $2H-MoS_2$, this intensity profiles would have no difference in contrast for the sample of more than two layers [1, 29]. Therefore, the sample shown in Figure 1 (i) is evident to be monolayer.

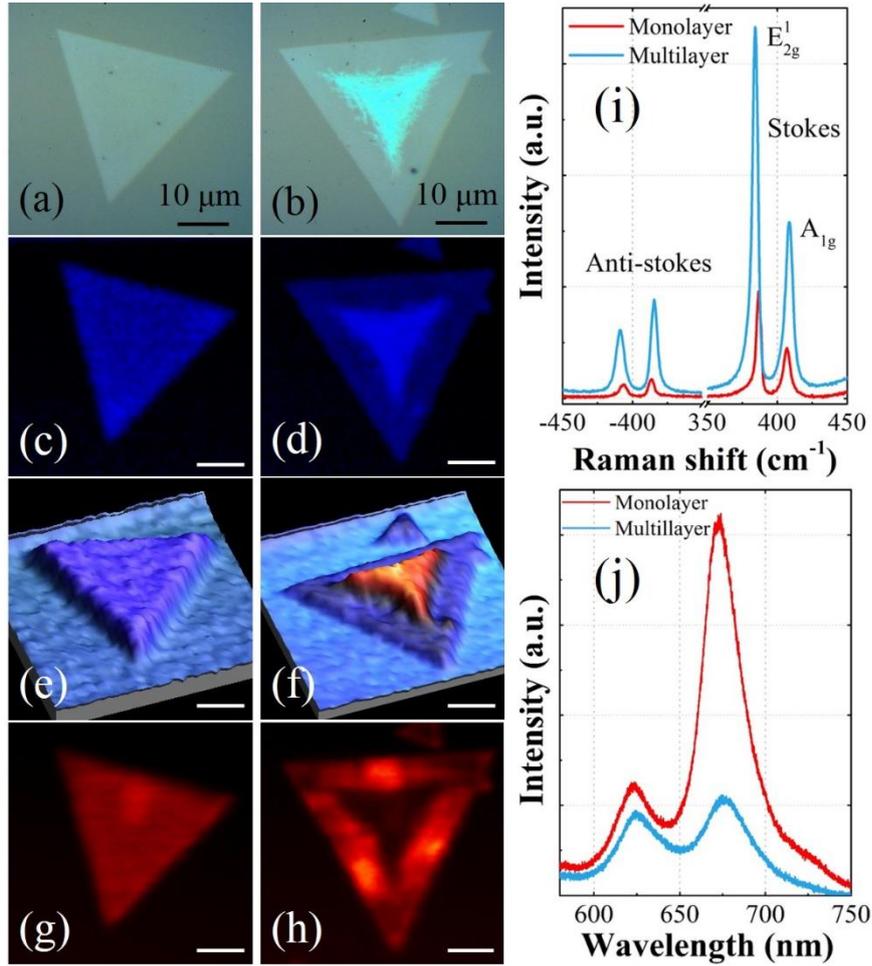

**Figure 2.** Optical microscopy image (100×) of (a) monolayer and (b) multilayer MoS$_2$ on quartz. (c) (d) E$_{2g}^1$ Raman mode mapping, (e) (f) 3D E$_{2g}^1$ Raman mode mapping, and (g) (h) PL mapping of the monolayer and multilayer MoS$_2$. The white scale bars equal to 10 μm. (i) Raman and (j) PL spectra of the monolayer and multilayer MoS$_2$.

Previous studies have pointed out that the frequency difference between two Raman peaks in MoS$_2$ is dependent on the layer number due to the anomalous lattice vibration of 2H MoS$_2$, which can be used to estimate the number of layers [30, 31]. In this work, Raman spectroscopy measurements of the MoS$_2$ on quartz were carried out by using a Monovista-P optical workstation (a confocal microscopy system) with a laser diode pumped laser at 532 nm. The Raman spectra of two triangular islands [Figure 2 (a) and (b)] were depicted in Figure 2 (i) and (j). The characteristic bands located at 386.7 and 406.6 cm$^{-1}$ [in red, from Figure 2 (a)] correspond to the in-plane ($E_{2g}^1$) and out-of-plane ($A_{1g}$) vibrational modes with the frequency difference of ~19.9 cm$^{-1}$, which illustrates that the island shown in Figure 2 (a) is monolayer. As for the blue curve from Figure 2 (b), the frequency difference increases to ~24.5 cm$^{-1}$, which is originated from the red shift of $E_{2g}^1$ and blue shift of $A_{1g}$ modes, and therefore the estimated layer number should be ~4 to 6. This is consistent with the AFM measurement. In addition, since the Raman spectrum of MoS$_2$ with 1T polytype does not show the $E_{2g}$ mode, it is evident that both the monolayer and multilayer in our work are 2H-MoS$_2$ with no evidence of structural distortion [1, 30, 32]. To obtain more detailed information about layer dependent Raman characteristics, we performed 2D Raman scan ($E_{2g}^1$ mode) over the two islands under same conditions (laser: 532 nm, 2 mW; exposure time: 1 s). As shown in Figure 2 (c), the homogeneous intensity distribution indicates the uniform and smooth surface of the monolayer island, whereas the central triangular part of the multilayer island is brighter because of more intense signal originating from thick layers of MoS$_2$ [Figure 2 (d)]. Additional visual images can be obtained by changing the mappings to three-dimensional (3D) views [Figure 2 (e) and (f)] that prominent regions represent higher Raman intensities. At the same time, anti-Stokes spectra were collected with two characteristic peaks located at -386.2 and -406.3 cm$^{-1}$ for the monolayer, -384.6 and -408.8 cm$^{-1}$ for the multilayer, respectively. The frequency difference increased from 20.1 cm$^{-1}$ to 24.2 cm$^{-1}$ along with the increase of layer thickness, which presents similar vibration behavior with Stokes.

Photoluminescence, which has been performed through the same confocal microscopy system, is another vital feature relevant with layer thickness [9]. As shown in Figure 2 (j), the two peaks at ~674 nm and 623 nm correspond to A (~1.9 eV) and B (~2.0 eV) direct excitonic transitions with the energy split from valence band spin-orbital coupling [10, 12-14]. Spatial intensity distributions of ~674 nm are shown in Figure 2 (g) and (h). The monolayer shows a dramatic improvement of PL efficiency compared with multilayer. It should be mentioned that bulk MoS$_2$ is not luminescent, and therefore the observed multilayer is not yet MoS$_2$ bulk [9].

## 2.2. Physical mechanisms of the distinct nonlinear absorption response in monolayer and multilayer MoS$_2$

As mentioned above, MoS$_2$ on transparent quartz can be used for nonlinear transmission/absorption experiment directly. A modified intensity-scan [33] system with microscopic imaging (μ-I-scan, see the Supporting Information), illustrated in Figure 3 (a), was used to investigate nonlinear absorption properties of the monolayer and multilayer MoS$_2$ on quartz. All experiments were performed by using 340 fs pulses from a mode-locked fiber laser operating at 1030 nm with the repetition rate of 1 kHz. As shown in Figure 3 (b), the monolayer MoS$_2$ exhibits a typical TPA effect as the normalized transmission reduces gradually with I$_0$ rising, and tends to be stable when I$_0$ above 60 GW/cm$^2$. Considering that the monolayer MoS$_2$ is a direct semiconductor with a band gap of ~1.89 eV [Figure 3 (d)], the electrons in valence band can absorb two photons simultaneously (1.2 eV for one photon) and transit to conduction band when excited by fs pulses at 1030 nm. If the incident pulses intensity is sufficiently large, the MoS$_2$ molecular population density in ground state (N$_1$) will decrease gradually, while the upper state (N$_2$) becomes populated due to this transition, which leads to the diminution of the TPA coefficient (β) and produces the degenerate TPA saturation effect.

Based on the NLO theory, the attenuation of a light beam [I (z)] passing through an optical medium caused by TPA can be described as [34]

$$\frac{dI(z)}{dz} = -\beta(I)I^2(z) \tag{1}$$

where β(I) is the TPA coefficient and z is the propagation distance in the sample. The linear absorption of the monolayer MoS$_2$ can be neglected because it does not satisfy the one photon absorption condition (1.2 eV < 1.89 eV). The change of TPA coefficient β over incident intensity (I$_0$) can be expressed as [34, 35]

$$\beta(I_0) = \frac{\beta_0}{1+(I_0/I_{s,\ TPA})^2} \qquad (2)$$

where $\beta_0$ is the nonsaturation TPA coefficient (a constant), and $I_{s,\ TPA}$ is the TPA induced saturable intensity of the monolayer $MoS_2$. In our experiment, we did not observe any clear nonlinear response from the quartz. As a result, the contribution of $\beta(I_0)$ only originates from the $MoS_2$ itself. As shown in Figure 3 (b), the results from Equation (1)-(2) fit the experimental data well with $\beta_0 \sim (7.62 \pm 0.15) \times 10^3$ cm/GW and TPA saturation intensity $I_{s,\ TPA} \sim (64.5 \pm 1.53)$ GW/cm$^2$. The on-focus beam radius is estimated to be ~17.3 μm, which is comparable to the OM images in consideration of the Gaussian distribution of the laser beam. As for comparison, we used the nonsaturation TPA model (normalized transmission: $T = 1/(1 + \beta_0' \cdot I_0 \cdot L)$) [34, 35] to fit the linear part of the T-$I_0$ curve (the black solid line in Figure 3 (b)). The result reveals $\beta_0' \sim (7.25 \pm 0.1) \times 10^3$ cm/GW, which is close to the result of TPA saturation model. It should be mentioned that $\beta(I_0)$ will decrease with the incident laser intensity increasing, which causes the experimental data diverging from the nonsaturation model curve, as shown in Figure 3 (b). However, we calculated the TPA coefficient at the damage threshold ($I_{th}$) of monolayer $MoS_2$ which is the minimum value of $\beta(I_0)$ within the scope of monoalyer $MoS_2$ damaging, and found the $\beta(I_{th})$ is as large as ~$(2.34 \pm 0.15) \times 10^3$ cm/GW. Although this vaule is less than the TPA coefficient of bilayer graphene (~$(20 \pm 4) \times 10^3$ cm/GW) [36] at ~1 μm, it is still larger than that of monolayer graphene [36] and the convenntional semiconductors, such as GaAs, GdS and ZnO [19-21]. That is to say, monolayer $MoS_2$ possess large TPA coefficient no matter at the low or higher excitation density before it is damaged. As reported in Ref. 23, the excitonic effect will prominently enhance the TPA coefficient in semiconductors. Cosidering monolayer $MoS_2$ has excitonic binding energy as large as ~0.96 eV for the lowest energy exciton [12] and that the TPA process belong to photoexcited excitonic absorption [22], it is likely that this large TPA coefficient is originated from the strong excitonic effect.

Furthermore, we calculated the TPA cross section ($\sigma_2$) of monolayer $MoS_2$ by using the equation [34]

$$\sigma_2 = \frac{\hbar\omega\beta_0}{N_{mono}} \quad (3)$$

where $N_{mono}$ is the $MoS_2$ molecule density (in units of cm$^{-3}$) in monolayer. The calculated $\sigma_2$ in our monolayer $MoS_2$ is ~$(4.45 \pm 0.09) \times 10^3$ GM (1 GM = $10^{-50}$ cm$^4$ s photon$^{-1}$ molecule$^{-1}$), with $N_{mono}$ ~$3.29 \times 10^{22}$ cm$^{-3}$, which is larger than certain organic molecules, such as rhodamine 6G and rhodamine B[37], but less than some porphyrin systems in femtosecond regime [34]. Notably, $MoS_2$ is an ionic crystal, and the $MoS_2$ molecule refers to the Mo and S atoms in each regular hexagon unit as shown in the inset of Figure 3 (d). Therefore, understandably, although the TPA coefficient of monolayer $MoS_2$ is giant and much greater than organic compounds, the TPA cross section is smaller than these materials' owing to the enormous $MoS_2$ molecule density. In addition, the damage threshold of the monolayer $MoS_2$ was confirmed to be as high as ~97 GW/cm$^2$ by real-time observation using the camera. The giant TPA absorption coefficient combined with the high damage threshold results in monolayer $MoS_2$ being a potential nanomaterials for photonic applications, such as optical limiter, optical beam shaper, etc.

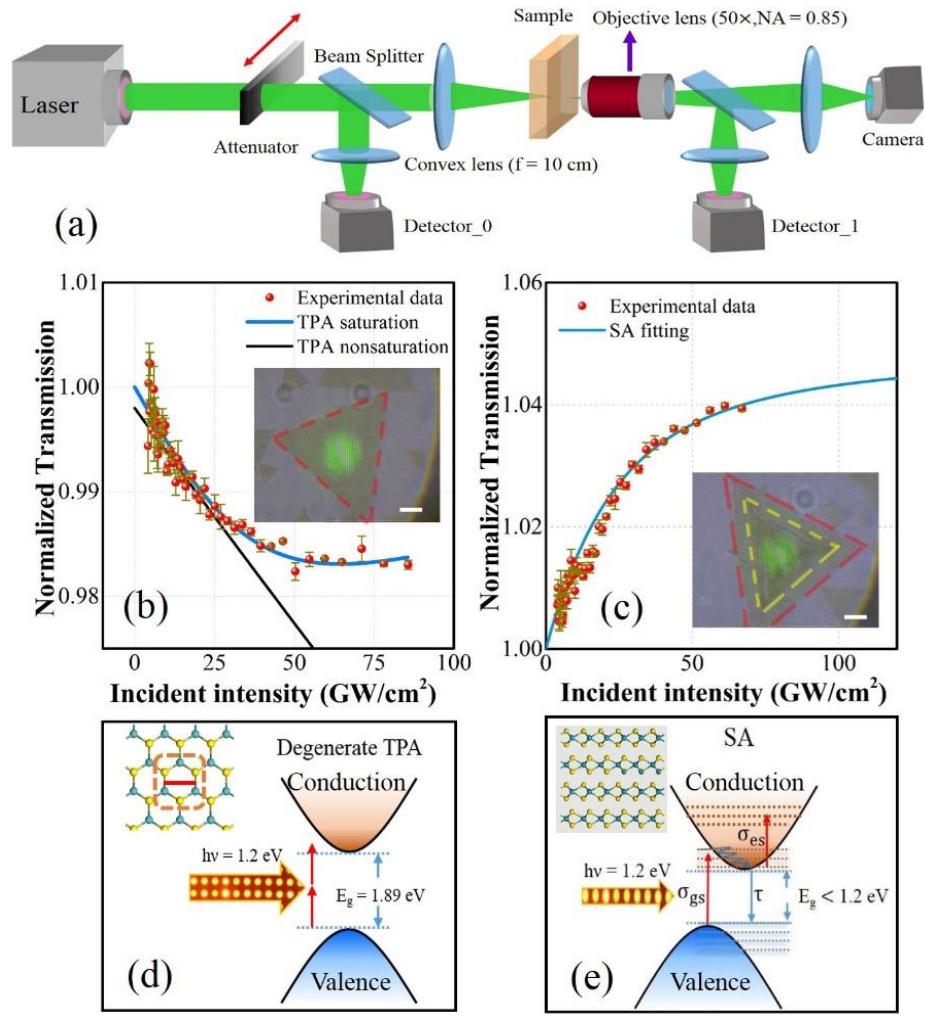

**Figure 3.** (a) Schematic diagram of the μ-I-scan system used for the nonlinear transmission/absorption experiment. (b) Degenerate TPA saturation effect of monolayer $MoS_2$, and (c) SA effect of multilayer $MoS_2$ under the excitation of 340 fs, 1030 nm, 1 kHz laser pulses. Inset: optical microscopy image of the laser spot irradiating on the monolayer and multilayer. (d) Degenerate TPA process of the monolayer $MoS_2$. Inset: Top-view of the monolayer $MoS_2$ with S-S (Mo-Mo) distance of 3.15-4.03 Å (red line),[29] and the dot line rectangle shows the regular hexagon unit containing one Mo atom and two S atoms. (e) Saturable absorption process of the multilayer $MoS_2$. Inset: the multilayer $MoS_2$ combined with weak van der Waals interaction between two adjacent layers.

In comparison with the monolayer, the multilayer MoS$_2$ shows a typical SA response under the same excitation conditions [Figure3 (c)]. Many previous studies have revealed that the few layer MoS$_2$ exhibits SA response even when excited by photons with energy less than the bulk MoS$_2$ bandgap 1.29 eV [2, 4-6]. Such a result is generally attributed to the band structure modulation caused by Mo or S atomic defects [2], or the coexistence of semiconducting and metallic states in the layered 2D materials [38]. For the excited photons ~1.2 eV (1030 nm) in our experiments, the multilayer MoS$_2$ also appears SA effect excited with one photon, which is in good agreement with the reported results [2, 3-6]. Figure 3 (d) depicts the band structure of the multilayer MoS$_2$ and a three-level model is employed to simulate the absorption process. The multilayer MoS$_2$ is considered as a slow saturable absorber, in which the excited state decay-time τ (100 ±10 ps [39]) is much longer compared to the pulse duration. By solving the rate equation, we can obtain the approximate analytical solution of the transmission T [40, 42]

$$T(L) = T_0 + \frac{T_{FN}(L) - T_0}{1 - T_0}(T_{max} - T_0) \qquad (4)$$

where $T_0 = e^{-N_{multi}\sigma_{gs}L}$ is the linear transmission in the low pulse intensity limit (94.1 % in our experiments), $N_{multi}$ is MoS$_2$ molecule density (in units of cm$^{-3}$) in multilayer, L is the sample thickness, and $\sigma_{gs}$ is the ground state absorption cross section. $T_{max} = e^{-N_{multi}\sigma_{es}L}$ is the high-energy (saturated) transmission limit achieved at very high pulse intensity, and $\sigma_{es}$ is the excited state absorption cross sections. The Frantz-Nodvik equation $T_{FN}(L)$ can be described as [40, 42]

$$T_{FN} = \frac{\ln\{1 + T_0[e^{\sigma_{gs}E(0)} - 1]\}}{\sigma_{gs}E(0)} \qquad (5)$$

where $E(0)$ is the incident beam fluence in units of photons per unit area. By fitting the experimental data using Equation (4)-(5), we determined $\sigma_{gs}$ and $\sigma_{es}$ of multilayer MoS$_2$ to be ~(8.77 ± 0.16) ×10$^{-17}$ cm$^2$ and (1.60 ± 0.03) × 10$^{-17}$ cm$^2$, respectively, with a corresponding absorber density ~1.4 ×10$^{21}$ cm$^{-3}$. The ratio of excited-state absorption cross section to ground-state cross section is deduced to be 0.18. In addition, the damage threshold of the multilayer

MoS$_2$ ~78 GW/cm$^2$ is lower than that of monolayer, which likely results from the extremely intense one-photon absorption in multilayer compared with the TPA in monolayer [18].

Overall, different nonlinear absorption properties between the pristine monolayer and multilayer MoS$_2$ were measured using the nonlinear transmission/absorption method and analyzed with the two nonlinear absorption models. To further clarify this difference, all the NLO parameters are summarized in Table 1.

**Table 1.** Nonlinear Absorption Parameters of MoS$_2$ in the fs Region

| CVD MoS$_2$ | Nonlinear absorption | Absorber density (cm$^{-3}$, ×10$^{22}$) | Nonsaturation absorption coefficient | Absorption cross section σ | I$_{sat}$ (GW/cm$^2$) | I$_{th}$ (GW/cm$^2$) |
|---|---|---|---|---|---|---|
| Monolayer | TPA | 3.29 | $\beta_0$ ~ (7.62 ± 0.15) × 10$^3$ cm/GW | $\sigma_2$ ~ (4.45 ± 0.09) × 10$^3$ GM | ~ 64.5 ± 1.53 | ~97 |
| Multilayer | SA | 0.14 | $\alpha_0$ ~ 8.69 × 10$^5$ cm$^{-1}$ | $\sigma_{gs}$ ~ (8.77 ± 0.16) × 10$^{-17}$ cm$^2$ | N.A. | ~78 |
|  |  |  |  | $\sigma_{es}$ ~ (1.60 ± 0.03) × 10$^{-17}$ cm$^2$ |  |  |

Finally, we observed the photoluminescence of the monolayer MoS$_2$ on SiO$_2$/Si under the excitation of 515 and 1030 nm fs laser with the repetition of 500 kHz, respectively. As shown in Figure 4, the blue curve depicts the frequency down-converted luminescence in monolayer MoS$_2$ pumped at 515 nm. The spectrum shows two characteristic bands located at ~670 and 623 nm corresponding to the A1 and B1 exciton direct transitions, respectively, which is consistent with the PL spectra shown in Figure 2 (j). More importantly, two-photon pumped frequency up-converted luminescence (the red spectral line) was directly observed with the characteristic band located at ~670 nm, which originates from the strong TPA in the monolayer MoS$_2$ at 1030 nm. The absence of the luminescence peak at ~623 nm possibly because the signal was too weak to detect. Notably, this frequency up-converted luminescence does not exist in multilayer MoS$_2$ due to the low quantum yield as well as the SA effect, so it is a vital nonlinear photoluminescence property to distinguish MoS$_2$ monolayer from the few-layer or bulk.

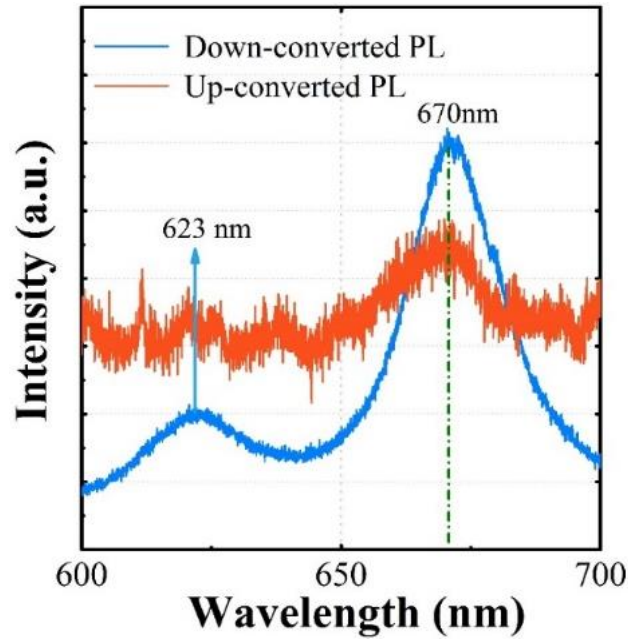

**Figure 4.** Frequency down-converted and up-converted PL of the monolayer MoS$_2$ on SiO$_2$/Si pumped by 515/1030 nm fs laser pulses.

## 3. Conclusion

In summary, monolayer and multilayer MoS$_2$ triangular islands have been synthesized on quartz and SiO$_2$/Si by seeding method via chemistry vapor deposition. Distinct NLO responses were demonstrated that multilayer MoS$_2$ showed SA effect. However, MoS$_2$ monolayer exhibited remarkable TPA effect for fs pulses at 1030 nm. The pure monolayer MoS$_2$ possessed a giant TPA coefficient of ~(7.62 ±0.15) ×10$^3$ cm/GW, which could originate from the strong excitonic effect in the monolayer MoS$_2$. This strong TPA effect combining with the high damage threshold of ~97 GW/cm$^2$ imply a potential application of this 2D nanomaterial for laser protection material in the NIR region. Multilayer MoS$_2$ showed one photon SA response with $\sigma_{es}/\sigma_{gs}$ ~0.18. Importantly, we have observed the direct two-photon pumped up-converted luminescence in the monolayer MoS$_2$ for the first time, which is another vital third order NLO response in 2D semiconductors.

**Acknowledgements** This work is supported in part by NSFC (No. 61178007 and 61308087), the External Cooperation Program of BIC, CAS (No. 181231KYSB20130007), Science and Technology Commission of Shanghai Municipality (No. 12ZR1451800) and the Excellent Academic Leader of Shanghai (No. 10XD1404600). J. Wang thanks the National 10000-Talent Program and CAS 100-Talent Program for financial support.

**References**

1. K. Wang, J. Wang, J. Fan, M. Lotya, A. O'Neill, D. Fox, Y. Feng, X. Zhang, B. Jiang, Q. Zhao, H. Zhang, J. N. Coleman, L. Zhang, and W. J. Blau, ACS Nano **7**, 9260-9267 (2013).

2. S. Wang, H. Yu, H. Zhang, A. Wang, M. Zhao, Y. Chen, L. Mei, and J. Wang, Adv. Mater. **26**, 3538-3544 (2014).

3. K. Wang, Y. Feng, C. Chang, J. Zhan, C. Wang, Q. Zhao, J. N. Coleman, L. Zhang, W. J. Blau, and J. Wang, Nanoscale **6**, 10530-10535 (2014).

4. H. Zhang, S. B. Lu, J. Zheng, J. Du, S. C. Wen, D. Y. Tang, and K. P. Loh, Opt. Express **22**, 7249-7260 (2014).

5. H. Liu, A. P. Luo, F. Z. Wang, R. Tang, M. Liu, Z. C. Luo, W. C. Xu, C. J. Zhao, and H. Zhang, Opt. Lett. **39**, 4591-4594 (2014).

6. J. Du, Q. K. Wang, G. B. Jiang, C. W. Xu, C. J. Zhao, Y. J. Xiang, Y. Chen, S. C. Wen, and H. Zhang, Sci. Rep. **4**, 6346 (2014).

7. D. Lagarde, L. Bouet, X. Marie, C. R. Zhu, B. L. Liu, T. Amand, P. H. Tan, and B. Urbaszek, Phys. Rev. Lett. **112**, 047401 (2014).

8. Q. H. Wang, K. Kalantar-Zadeh, A. Kis, J. N. Coleman, and M. S. Strano, Nat. Nanotechnol. **7**, 699-712 (2012).

9. A. Splendiani, L. Sun, Y. Zhang, T. Li, J. Kim, C. Y. Chim, G. Galli, and F. Wang, Nano lett. **10**, 1271-1275 (2010).

10. D. Xiao, G. B. Liu, W. Feng, X. Xu, and W. Yao, Phys. Rev. Lett. **108**, 196802 (2012).

11. M. Xu, T. Liang, M. Shi, and H. Chen, Chem. Rev. **113**, 3766-3798 (2013).

12. D. Y. Qiu, F. H. da Jornada, and S. G. Louie, Phys. Rev. Lett. **111**, 216805 (2013)

13. A. Ramasubramaniam, Phys. Rev. B **86**, 115409 (2012).

14. T. Cheiwchanchamnangij, and W. R. L. Lambrecht, Phys. Rev. B **85**, 205302 (2012).

15. X. Yin, Z. Ye, D. A. Chenet, Y. Ye, K. O'Brien, J. C. Hone, and X. Zhang, Science **344**, 488-490 (2014).

16. L. M. Malard, T. V. Alencar, A. P. M. Barboza, K. F. Mak, and A. M. de Paula, Phys. Rev. B **87**, 201401 (2013).

17. R. Wang, H. C. Chien, J. Kumar, N. Kumar, H. Y. Chiu, and H. Zhao, ACS Appl. Mater. Inter. **6**, 314-318 (2014).

18. I. Paradisanos, E. Kymakis, C. Fotakis, G. Kioseoglou, and E. Stratakis, Appl. Phys. Lett. **105**, 041108 (2014)

19. T. D. Krauss, and F. W. Wise, Appl. Phys. Lett. **65**, 1739-1741 (1994).

20. J. Lami, P. Gilliot, and C. Hirlimann, Phys. Rev. Lett. **77**, 1632-1635 (1996).

21. C. M. Cirloganu, L. A. Padilha, D. A. Fishman, S. Webster, D. J. Hagan, and E. W. Van Stryland, Opt. Express **19**, 22951-22960 (2011).

22. M. Sheik-Bahae, D. J. Hagan, and E. W. Van Stryland, Phys. Rev. Lett. 65, 96-99 (1990).

23. E. W. Van Stryland, M. A. Woodall, H. Vanherzeele, and M. J. Soileau, Opt. Lett. **10**, 490-492 (1985).

24. X. Ling, Y. H. Lee, Y. Lin, W. Fang, L. Yu, M. S. Dresselhaus, and J. Kong, Nano lett. **14**, 464-472 (2014).

25. Y. H. Lee, L. Yu, H. Wang, W. Fang, X. Ling, Y. Shi, C. T. Lin, J. K. Huang, M. T. Chang, C. S. Chang, M. Dresselhaus, T. Palacios, L. J. Li, and J. Kong, Nano lett. **13**, 1852-1857 (2013).

26. Y. H. Lee, X. Q. Zhang, W. Zhang, M. T. Chang, C. T. Lin, K. D. Chang, Y. C. Yu, J. T. Wang, C. S. Chang, L. J. Li, and T. W. Lin, Adv. Mater. **24**, 2320-2325 (2012).

27. A. M. Van der Zande, P. Y. Huang, D. A. Chenet, T. C. Berkelbach, Y. You, G. H. Lee, T. F. Heinz, D. R. Reichman, D. A. Muller, and J. C. Hone, Nat. Mater. **12**, 554-561 (2013).

28. S. Najmaei, Z. Liu, W. Zhou, X. Zou, G. Shi, S. Lei, B. I. Yakobson, J. C. Idrobo, P. M. Ajayan, and J. Lou, Nat. Mater. **12**, 754-759 (2013).

29. J. N. Coleman, M. Lotya, A. O'Neill, S. D. Bergin, P. J. King, U. Khan, K. Young, A. De, S. Gaucher, R. J. Smith, I. V. Shvets, S. K. Arora, G. Stanton, H. Y. Kim, K. Lee, G. T.

Kim, G. S. Duesberg, T. Hallam, J. J. Boland, J. J. Wang, J. F. Donegan, J. C. Grunlan, G. Moriarty, A. Shmeliov, R. J. Nicholls, J. M. Perkins, E. M. Grieveson, K. Theuwissen, D. W. McComb, P. D. Nellist, and V. Nicolosi, Science **331**, 568-571 (2011).

30. C. Lee, H. Yan, L. E. Brus, T. F. Heinz, J. Hone, and S. Ryu, ACS Nano **4**, 2695-2700 (2010).

31. H. Li, Q. Zhang, C. C. R. Yap, B. K. Tay, T. H. T. Edwin, A. Olivier, and D. Baillargeat, Adv. Funct. Mater. **22**, 1385-1390 (2012).

32. D. Yang, S. J. Sandoval, W. M. R. Divigalpitiya, J. C. Irwin, and R. F. Frindt, Phys. Rev. B **43**, 12053 (1991).

33. B. Taheri, H. M. Liu, B. Jassemnejad, D. Appling, R. C. Powell, and J. J. Song, Appl. Phys. Lett. **68**, 1317-1319 (1996).

34. G. S. He, L. S. Tan, Q. Zheng, and P. N. Prasad, Chem. Rev. **108**, 1245-1330 (2008).

35. R. Schroeder, and B. Ullrich, Opt. Lett. **27**, 1285-1287 (2002).

36. H. Yang, X. B. Feng, Q. Wang, H. Huang, W. Chen, A. T. S. Wee, and Wei Ji, Nano lett. **11**, 2622-2627 (2011).

37. N. S. Makarov, M. Drobizhev, and A. Rebane, Opt. Express **16**, 4029-4047 (2008).

38. J. Zheng, H. Zhang, S. Dong, Y. Liu, C. T. Nai, H. S. Shin, H. Y. Jeong, B. Liu, and K. P. Loh, Nat. Commun. **5**, 2995 (2014).

39. R. Wang, B. A. Ruzicka, N. Kumar, M. Z. Bellus, H. Y. Chiu, and H. Zhao, Phys. Rev. B **86**, 045406 (2012).

40. Z. Burshtein, P. Blau, Y. Kalisky, Y. Shimony, and M. R. Kikta, IEEE J. Quantum Elect., **34**, 292-299 (1998).

41. W. Wang, Y. Bando, C. Zhi, W. Fu, E. Wang, and D. Golberg, J. Am. Chem. Soc. **130**, 8144-8145 (2008).

42. X. Y. Zhang, S. F. Zhang, C. X. Chang, Y. Y. Feng, Y. X. Li, N. N. Dong, K. P. Wang, L. Zhang, W. J. Blaub, and J. Wang, Nanoscale **7**, 2978-2986 (2015).

Supporting Information

**Giant Two-Photon Absorption Coefficient and Frequency Up-Converted Luminescence in Monolayer MoS$_2$**

*Yuanxin Li, Ningning Dong, Saifeng Zhang, Xiaoyan Zhang, Yanyan Feng, Kangpeng Wang, Jun Wang* [*]

*Corresponding Author: jwang@siom.ac.cn

*MoS$_2$ preparation:* The MoS$_2$ monolayer and multilayer islands were synthesized by seeding method via CVD growth. The PTAS seeds solution was prepared by alkaline hydrolysis of 3, 4, 9, 10 - perylene tetracarboxylic dianhydride (PTCDA) with a concentration of 50 mg/mL [41]. Subsequently, a drop of PTAS solution was spin coated on substrates at 1000 r/min for 1.5 minutes. The substrates were strictly cleaned in acetone and piranha solution for 30 minutes, and dried in vacuum drying oven for 1h before used for spin coating. The deposited substrates were placed face down on a ceramic boat containing 10 mg MoO$_3$ (≥99.98% Sigma Aldrich) and loaded into a two-inch CVD furnace. Another boat containing 30 mg S (≥99.998% Sigma Aldrich) was settled in the front of the quartz tube, and the distance between the two boats was ~20 cm. The CVD growth was performed at atmospheric pressure with ultrahigh-purity Argon as carrier gas. The furnace temperature was programmed to rise from room temperature to 101℃ in 10 min with 200 s.c.c.m. Ar, sit 1h at 101℃; ramp to the growth temperature 650℃ at 15℃ min$^{-1}$ with 50 s.c.c.m. and sit 3 min; cool down naturally to 550℃ with 50 s.c.c.m., and open furnace for rapid cooling with flowing 500 s.c.c.m. Ar.

*μ-I-scan:* In this system, the objective lens and camera were used to observe the samples, and check whether the laser spot was tightly focused and exactly irradiated at the island center. As shown in the insets of Figs. 3 (b) and (c), the laser spot diameter was ~20 μm, so the laser pulses could pass through the islands completely. Two high-accuracy photoelectric detectors (detector_0 and detector_1) were used to record the referenced and transmitted laser pulse energy, and the ratio of detector_1 to detector_0 was defined as transmission (T). An attenuator driven by a fine linear translation stage was used to change the incident light intensity (I$_0$), thus the dependence of transmission on I$_0$ can be obtained (T-I$_0$ curve).